\documentclass{PoS}
\title{Angular Momentum Loss by Magnetic Braking and Gravitational Radiation in Relativistic Binary Stars}
\ShortTitle{Angular Momentum Loss in Relativistic Binary Stars}

\author{{K. Yakut}\\
        University of Cambridge, Institute of Astronomy, Madingley Road, Cambridge CB3 0HA, UK\\
        University of Ege, Department of Astronomy \& Space Sciences, 35100  \.Izmir,  Turkey \\
        E-mail: \email{yakut@ast.cam.ac.uk}}

\author{B. Kalomeni\\
        \.Izmir Institute of Technology, Department of Physics, 35430 \.Izmir, Turkey \\
        E-mail: \email{belindakalomeni@iyte.edu.tr}}

\author{C. A. Tout\\
        University of Cambridge, Institute of Astronomy, Madingley Road, Cambridge CB3 0HA, UK\\
        E-mail: \email{cat@ast.cam.ac.uk}}
\abstract{Angular momentum loss (AML) mechanisms and dynamical evolution owing to magnetic braking and
gravitational radiation in relativistic binary stars (RBS) are studied with use of physical parameters collected from the literature.
We have calculated and compared AML time scales for the RBS with non-degenerate components and double degenerate (DD) systems.}

\FullConference{VII Microquasar Workshop: Microquasars and Beyond\\
		 September 1-5 2008\\
		 Fo\c{c}a, \.Izmir, Turkey}

\begin{document}

\section{Introduction}
White dwarfs (WD), neutron stars (NS), and black holes (BH) represent the endpoint of stellar evolution.
Compact objects provide conditions to study physics at high energy, magnetic fields, temperatures etc.,
and they are dominated by gravitation. Important properties of these compact objects can be
determined if they are  members of binary systems. This make possible to determine the orbital
elements and physical parameters from  precise data obtained by various techniques.
Such studies are important to test the predictions of stellar evolution theory.
Additionally, binary systems with a compact companion or double degenerate (DD), owing to their general
relativistic effects, are good laboratories to test general relativity.

The literature has been scanned for the physical parameters of High Mass X-Ray Binary (HMXRB)
and Low Mass X-Ray Binary (LMXRB) systems with a black hole (BH) or neutron star (NS) companion,
as well as other systems listed in Table~\ref{table-relbin}.
We use the data obtained to study the angular momentum loss mechanisms and focus on the
dynamical evolution driven by magnetic braking and gravitational radiation in relativistic binaries.

\section{Data selection}
In modern astrophysics, theoretical and observational studies of compact objects, with both
ground based and satellite observations, offer potential to research the formation and evolution.
In recent years, as a result of the advanced technology and instrumentation,
the number of papers on relativistic binaries has increased. These studies contain significantly
accurate orbital elements and physical parameters. The mass interval of these binary systems
are summarized in Table~1. Isolated black holes, neutron
stars and white dwarfs are included for comparison. Using these parameters
collected from the literature we have studied their possible angular momentum loss mechanisms.

\section{Angular Momentum Loss}
The orbital period of the binary system can be determined accurately.
Any orbital period variation in tidally locked binary systems is due to
evolutionary changes or the presence of a third body orbiting the binary.
In close binaries, in general, we can estimate the angular momentum variation via
observations of period changes. Angular momentum loss leads to a period decrease.
It can be studied as the orbital angular momentum and spin angular momentum.
The spin angular momentum (H$^{\rm{s}}$) for a single star is given in the form,
\begin{equation}
H_{\rm{i}}^{\rm{s}}=I\omega=k^2R_i^2M_i\omega_i, \label{Eq-hspin}
\end{equation}
where, subscript $i$ stands for the components, $i=1$ for the primary  and $i=2$ for the secondary, respectively,
$I$ is the star's moment of inertia, $\omega$ is the spin angular velocity, $k^2$ is the gyroton radius,
$R_i$ is the radius of the star, and $M_i$ is the mass of the component.
The total orbital angular momentum is
\begin{equation}
{H_{\rm{t}}^{\rm{o}}} ={M_1 M_2}\sqrt{\frac{G a (1-e^2)}{M}} \\
  =111 q (1+q)^{-2} (\frac{M_t}{M_\odot})^{-5/3} (\frac{P}{\rm{day}})^{1/3} \sqrt{1-e^2} \,\,(M_\odot R_\odot ^2 /{\rm{day}})\label{Eq-AMosolar}
\end{equation}
 where  $G$ is gravitational constant, $a$  is the separation of two stars and $q$ is the mass ratio of the components specify $q=\frac{M_2}{M_1}$.

The angular momentum is extracted from the binary orbit because of the
\emph{(i)} magnetic braking in a magnetized stellar wind (MSW), \emph{(ii)} gravitational radiation (GR),
\emph{(iii)} non-conservative mass transfer and \emph{(iv)} a third (or multi-) body effect.
The total system angular momentum loss is the sum of all these effects.
We assume throughout this study that relativistic binary star evolution is driven by GR and MSW.
\setcounter{table}{0}
\begin{table*}
\begin{center}
\scriptsize
\caption[]{Mass distribution  of relativistic binaries and isolated compact stars based observational data set.}\label{table-relbin}
\begin{tabular}{lllllll}
\hline
	&		&component		&	M$_1$	&		&	M$_2$ 	&		\\
Type	&	pri+sec	&	n	&	max.	&	min.	&	max.	&	min.	\\
\hline
HMXRB 	&	BH+S	&	6	&	23.1	&	6.0	&	70	&	6.5	\\
HMXRB 	&	NS+S	&	11	&	2.4	&	1.0	&	58	&	8.8	\\
LMXRB 	&	BH+S	&	13	&	14	&	4.0	&	2.7	&	0.37	\\
LMXRB 	&	NS+S	&	6	&	1.8	&	1.4	&	2.3	&	0.4	\\
NLCV	&	WD+S/B	&	39	&	1.4	&	0.3	&	1.1	&	0.05	\\
mCV	    &	WD+S	&	10	&	0.8	&	0.4	&	0.5	&	0.1	\\
PreCV 	&	WD+S	&	21	&	0.84	&	0.39	&	0.93	&	0.1	\\
DDNSNS	&	NS+NS	&	8	&	1.6	&	1.14	&	1.4	&	1.05	\\
DDNSWD	&	NS+WD	&	9	&	2.1	&	1.27	&	1.3	&	0.16	\\
DDWDWD	&	WD+WD	&	10	&	0.7	&	0.32	&	0.7	&	0.29	\\
AMCVn	&	WD+WD	&	5	&	0.98	&	0.59	&	0.13	&	0.011	\\
\hline
Black Hole	   &	 	&	19	&	23.1	&	4.0	&	 	&	 	\\
Neutron Star   &	 	&	42	&	2.4	&	1.0	&	 	&	 	\\
White Dwarf	   &	 	&	90	&	1.4	&	0.01	&	 	&	 	\\
\hline
\end{tabular}
\end{center}
\end{table*}
\subsection{Magnetic Braking in a Magnetized Stellar Wind}
The Sun is known to rotate more rapidly at the equator than at the poles.
The speed of this differential rotational and convection are important
for the production of magnetic fields in the convective zone (see Yakut \& Eggleton and references therein).
Light curves of close binary stars show  one maximum  higher than the other.
This phenomenon is called the O'Connell effect (Milone 1968).
These asymmetries are usually attributed to the magnetic activity, though the effect is
 not satisfactorily explained by existing the models.
A large number of late-type stars show magnetic activity indicators.
Hence, activity may play an important role in relativistic binaries whose components
are late-type stars (M $\lesssim$ 1.5 M$_\odot$) .

The Skumanich relation (Skumanich 1972, Smith 1979 ) predicts a strong correlation between equatorial rotation velocity and age as
$$v_{\rm{e}} = 10^{14} \times f \times t_{\rm{0}}^{-0.5}$$ cm s$^{-1}$, where $t_{\rm{0}}$ is the age of
the star and the value of $f$ was determined empirically by Skumanich (1972) to be 0.7 and Smith (1979) to be 1.78.
The Skumanich law and Eq.~(\ref{Eq-hspin}) yield the angular momentum loss by MSW,
\begin{equation}
\left(\frac{dH}{dt}\right) _{\textrm{MSW}}=-1.6\times 10^{-30}M_2R_2^{4}\omega^3 .\label{Eq-aml-msw}
\end{equation}
The angular momentum loss time scale
\begin{equation}
\tau_{MSW} = 14 (\frac{M}{M_\odot})^{2/3} (\frac{R_2}{R_\odot})^{-4}(\frac{P}{\rm{day}})^{10/3}(1+q)^{-1}(1-e^2)^{1/2} \times \rm{Gyr}.\label{Eq-timescale-msw}
\end{equation}

If the secondary stars in binaries fill their Roche lobes then the radius may be  calculated with the relation given by Eggleton (1983),
\begin{equation}
R_2=R_{\rm L}=a\frac{0.49q^{2/3}}{0.6q^{2/3}+\ln
(1+q^{1/3})}.
\end{equation}
Otherwise we used the mass-radius relations
\begin{equation}
R=M^{0.97} \,\,\,\,\,\,
\textrm{for} \,\,\,\,\,  M_2\leq 0.5 M_\odot \,\,\,\,\,\,
\textrm{and} \,\,\,\,\,\,
R=M^{1.11} \,\,\,\,\textrm{for}\,\,\,\,\,  0.5 M_\odot < M_2\leq 1.5 M_\odot
\end{equation}
given by Yakut et al. (2008a).

\subsection{Gravitational Radiation}
Binary star systems, in which both components are white dwarfs, neutron stars,
or black holes emit gravitational waves that can carry off angular momentum.
The shape of the binary orbit and the energy of the system given by
\begin{equation}
r=\frac{a(e^2-1)}{1+e \cos \nu} \,\,\,\,\,\,\,\,\,\,\,\,
\textrm{and} \,\,\,\,
E=\frac{-GM_1M_2}{2a}.
\end{equation}
These equations and the semi-major axis variation rates (Landau \& Lifshitz, 1951) yield
the angular momentum loss by the gravitational radiation and the angular momentum loss time scale (in Gyr)
\begin{equation}
\frac{da}{dt}=  - \frac{64}{5}\frac{G^3}{c^5} \frac{M_1M_2}{a^3} (M_1+M_2), \label{Eq-da}
\end{equation}

\begin{equation}
\left(\frac{dH}{dt}\right)_{GR}= -3.44\times10^{44}(M_1M_2M^{-1/3})^2 P^{-7/3} f(e) \label{Eq-AMLGR}
\end{equation}

\begin{equation}
\tau_{GR} = 376.4q^{-1}(1+q)^2M^{-5/3}P^{8/3} (1-e^2)^{3/2}(1+\frac{7}{4}e^2)^{-1}  \textrm{Gyr}. \label{Eq-timescale-gr}
\end{equation}

\subsection{Timescales}
The angular momentum loss time scales for relativistic and non-relativistic
 binary stars (LTCB and HTCB) are shown in Figs.1 and 2. The details will be discussed in Yakut et al. (2008b).
 Fig. 3 is the same as Figs.1 and 2 but for double compact degenerate binary systems.

\begin{figure}
\includegraphics[width=.7\textwidth]{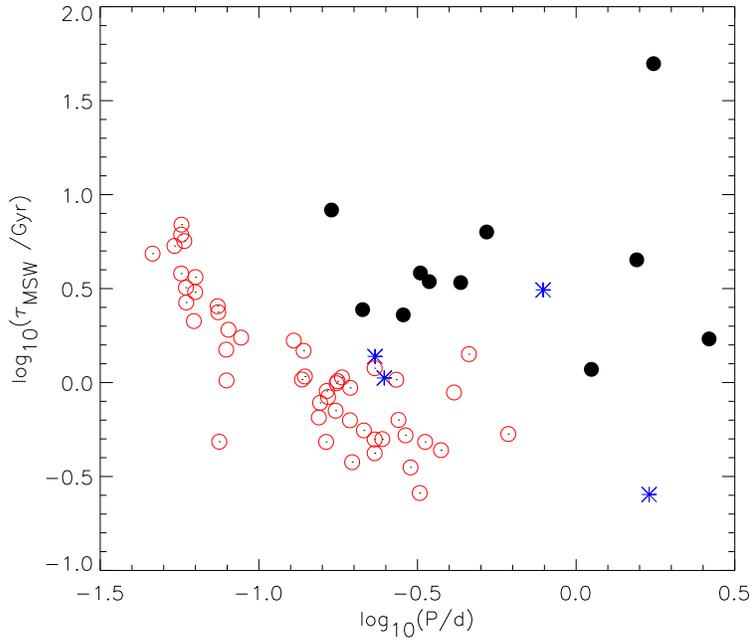}
\caption{Plot of  $\log \tau_{\textrm{MSW}}$ {\it vs.} $\log$P. The open circles, asterisks, and filled circles show binaries with WD, NS and BH companions, respectively. }
\label{fig1}
\end{figure}

\begin{figure*}\includegraphics[width=.7\textwidth]{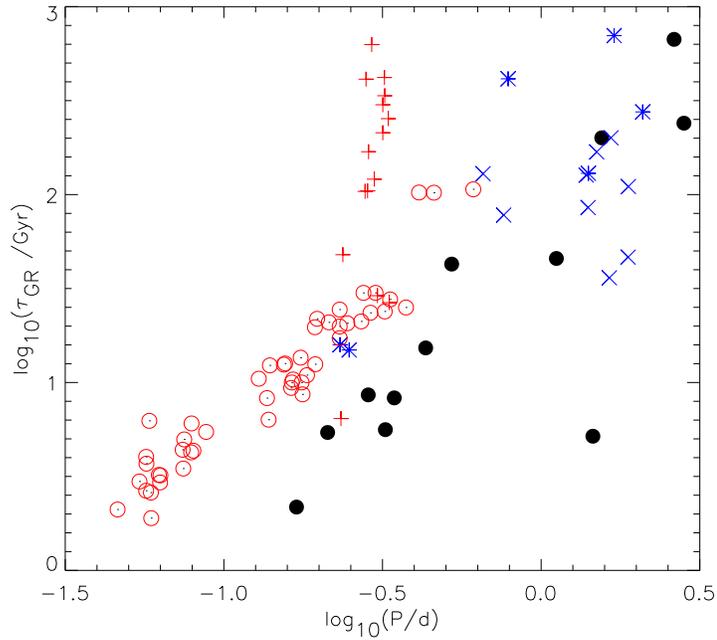}
\caption{Plot of  $\log \tau_{\textrm{GR}}$ {\it vs.} $\log$P. The open circles, asterisks, filled circles show
binaries with WD, NS, and BH companions and + and x show the LTCB  and HTCB systems, respectively.}\label{fig2}
\end{figure*}

\begin{figure*}\includegraphics[width=.7\textwidth]{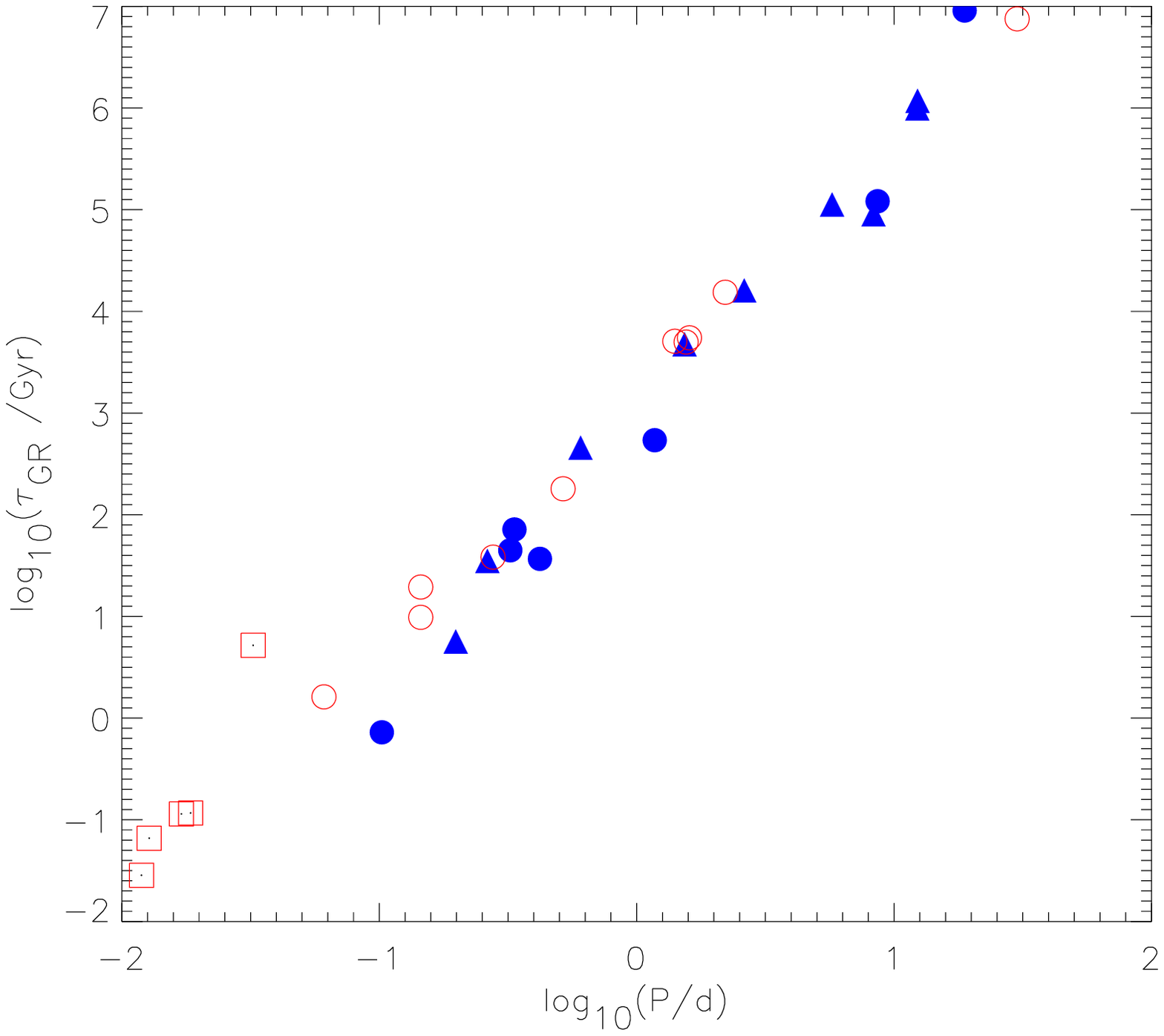}
\caption{Plot of  $\log \tau_{\textrm{GR}}$ {\it vs.} $\log$P  for double degenerate binaries.
The open circles, open squares, filled circles and triangles indicate
WD+WD, WD+WD (AMCVn), NS+NS and NS+WD binaries, respectively. }\label{fig3}
\end{figure*}

\section{Conclusion}
The aim is to collate the physical properties of the relativistic binary stars because they are
valuable in studies of formation and evolution of compact objects. From observational data (see Table~1)
we also summarize mass distribution for the compact objects. According to the data set, maximum value for
mass of black holes, neutron stars and white dwarfs are 23, 2.4 and 1.4 $M_\odot$, respectively and for minimum mass
for black holes, neutron stars and white dwarfs are 4, 1, and 0.01 respectively. We also draw attention to gap
between minimum black hole masses and maximum mass of neutron stars (see for details Yakut et al. 2008b).
The data obtained have been used to study the angular momentum loss and its evolution driven by magnetic braking and
gravitational radiation. Binary stars with BH components have longer orbital periods them those
with WD and NS components. Increase in the orbital period increases the efficiency of AML by GR.
Greater the difference between the mass ratio the shorter is the AML time-scale via MSW.

\textbf{Acknowledgments}\\
KY and BK acknowledges support by the Turkish Scientific and Technical Research Council (T\"UB\.ITAK).

\end{document}